\documentclass[
    ,final            % use final for the camera ready runs
  ]
  {aipproc}
\layoutstyle{6x9}
\usepackage{subfigure}
\usepackage{mathrsfs}
\usepackage{graphicx}
\usepackage{times}
\usepackage{pgf,pgfarrows,pgfnodes,pgfautomata,pgfheaps,pgfshade}
\usepackage{amsmath,amssymb}
\usepackage{fix-cm}
\usepackage{geometry}
\usepackage[cmtip,arrow]{xy}
\usepackage{pb-diagram,pb-xy}
\usepackage{rotating}

\newcommand{\beq}{\begin{equation}}
\newcommand{\eeq}{\end{equation}}

\input{aipcheck}
\begin{document}

\title{Non-relativistic Extended Gravity and its applications 
across different astrophysical scales}
\author{J.~C.~Hidalgo}{address = {Instituto de Astronom\'{\i}a,
Universidad Nacional Aut\'onoma de M\'exico, AP 70-264, Distrito
Federal 04510, M\'exico.} } 
\author{S.~Mendoza}{address = {Instituto de Astronom\'{\i}a,
Universidad Nacional Aut\'onoma de M\'exico, AP 70-264, Distrito
Federal 04510, M\'exico.} } 
\author{X.~Hernandez}{address = {Instituto de Astronom\'{\i}a,
Universidad Nacional Aut\'onoma de M\'exico, AP 70-264, Distrito
Federal 04510, M\'exico.} } 
\author{T. Bernal}{address = {Instituto de Astronom\'{\i}a,
Universidad Nacional Aut\'onoma de M\'exico, AP 70-264, Distrito
Federal 04510, M\'exico.} } 
\author{M.~A.~Jimenez}{address = {Instituto de Astronom\'{\i}a,
Universidad Nacional Aut\'onoma de M\'exico, AP 70-264, Distrito
Federal 04510, M\'exico.} } 
\author{C.~Allen}{address = {Instituto de Astronom\'{\i}a, Universidad
Nacional Aut\'onoma de M\'exico, AP 70-264, Distrito Federal 04510,
M\'exico.} } 
\date{\today}

\begin{abstract}
Using dimensional analysis techniques we present an extension
of Newton's gravitational theory built under the assumption that
Milgrom's acceleration constant is a fundamental quantity of nature.
The gravitational force converges to Newton's gravity and to a MOND-like
description in two different mass and length regimes.  It is shown that
a modification on the force sector (and not in the dynamical one as MOND
does) is more convenient and can reproduce and predict different
phenomena usually ascribed to dark matter at the non-relativistic
level.  
\end{abstract}
\maketitle

%%%%%%%%%%%%%%%%%%%%%%%%%%%%%%%%%%%%%%%%%%%%
%% MAINMATTER
%%%%%%%%%%%%%%%%%%%%%%%%%%%%%%%%%%%%%%%%%%%%

\section{Extended Gravity}

  The gravitational anomalies in the dynamics of galaxies and larger 
systems are the basis to expect non-baryonic dark matter.
Since gravitational experiments have been performed in systems no larger
than the mass and length scales associated to those of the Solar System, 
then  it is necessary to explore an extension of the gravitational theory
avoiding an elusive dark matter component.

Milgrom \cite{milgrom:80} developed a modification in the dynamical
sector of Newton's second law in such a way that, when a test particle
experiences a gravitational acceleration greater than an acceleration
constant \(a_0 \sim 10^{-10}\textrm{m}/\textrm{s}^2 \), a
Newtonian behaviour is expected and, when the inequality is inverted, the
acceleration exerted is proportional to the square root of the
Newtonian value. A relativistic Tensor-Vector-Scalar (TeVeS
\cite{bekenstein:04}) generalisation of this has been proposed, but
its inconsistency with gravitational lensing \cite{mavromatos:09} and the
null-observations by the terrestrial Newtonian law of inertia
experiments indicate that the modification has to be done in
the gravitational sector and not in the dynamical one. 

  Here we assume that Milgrom's acceleration constant \(a_0\) is of
fundamental character. The acceleration \( a \) felt by a test
particle at a distance \( r \) from the mass \( M \) generating the
gravitational field must also be a function of Newton's constant of gravity
\( G \), therefore, Buckingham's theorem of dimensional analysis demands that
\cite{mendoza:10}:
\begin{equation}
  a = a_0 f(x) \rightarrow a_0 
    \begin{cases}
       x := l_M / r := \left( G M a_0^{-1}  \right)^{1/2} / r, \qquad
       \text{for \( x\ll 1 \) },  \\
       x^2 = ( l_M / r )^2 =  G M a_0^{-1} / r^2,  \qquad \text{for \( x
       \gg 1 \)}.
    \end{cases}
\label{limits}
\end{equation}

\noindent As explained by \cite{mendoza:10}, where the theory first
appeared, the introduction of a mass-length scale \( l_M \)  breaks
the scale-invariance of gravity.  This Extended Newtonian theory of
gravity (ENG) modifies the gravitational force and leaves the inertia
law intact.  The force law is constructed from expansions in Taylor
series about the limits~\eqref{limits}, obtaining \cite{mendoza:10}: 
\begin{equation}
  \label{extended:gravity}
    {a}/{a_0} = f(x) = x ({1 - x^{n + 1}})/ ({1 - x^n}), 
\end{equation}

\noindent where the parameter $n$  modulates
the smoothness of the transition between the Newtonian and the MONDian
regimes. In fact, a fit to the rotation profile
of the Milky Way indicates that $n = 3$  \cite{mendoza:10}. Also,
Newton's theorems remain valid on spherically symmetric matter
distributions. 

\section{Tests and probes across astrophysical scales}

A few testable and falsifiable predictions can be derived in the ENG
since the experimental confirmation of Newtonian gravity for $x
< 1$ would discard the theory completely. Conversely, the
theory would not be an alternative if observations infer the
existence of dark matter at scales above $x=1$.  In regard of this, we
have explored environments to probe gravity at the non-relativistic level.  
 
In \cite{mendoza:10}, the predictions of ENG were tested over several
astrophysical scales. In our Solar System,
departures from the Newtonian acceleration~\eqref{extended:gravity}
lie within the uncertainty of the orbits of planets.  Recently
\cite{exirifard:11} has shown the same trend for the measured acceleration
of the Moon's orbit.

  In ENG, a gravitational system in mechanical equilibrium has a velocity
dispersion \( \sigma = \left( G M a_0 \right)^{1/4} g(x) \) according
to Buckingham's theorem of dimensional analysis.  A particular class
of astrophysical objects is characterised by a mass \( M \) and length \( r \)
and so,  the general function \( g(x) \) can be approximated
by a power-law \( g(x) = x^{\alpha} \), i.e.
\begin{equation}
  \sigma(r) \propto r^\beta M^\gamma, \quad \text{where} \quad
  \beta = 1/2 - 2 \gamma =
                \begin{cases}
                  -1/2 & \text{if\, { $x \gg 1$;}}\\
                  0    &\text{if\, { $x \ll 1$.}}
                \end{cases}
\label{obs:law}
\end{equation}

\noindent  with transition values between these two limits for systems
with $x \approx 1$. The above equation reproduces accurately the observed
scaling laws in a variety of systems \cite{hernandez:09,mendoza:10}.
Namely, it converges to the Faber-Jackson relation for elliptical galaxies
when \( x\ll 1 \). In the same limit, allowing for a proportionality
between isotropic velocity dispersions in pressure supported systems
and rotation velocities in angular momentum supported ones,
Eq.~\eqref{obs:law} reproduces the baryonic Tully-Fisher relation.  When \(
x \gg 1 \), this equation is the virial equilibrium relation of
Newtonian gravity \( r \propto \sigma / \sqrt{G \rho} \), where \(
\rho \) is the characteristic mass density of the system.  
Consequently, the galactic Faber-Jackson relation in the MONDian regime
appears as the Jeans equilibrium relation in the Newtonian one.  For
intermediate scales \( x \sim 1 \), the equilibrium for local
dwarf Spheroidals is also reproduced. Finally, the Fundamental plane of
elliptical galaxies, for which both indices \( \beta \) and \( 
\gamma \) are directly measured, is consistent with ENG.

\begin{figure*}
 \begin{minipage}[b]{7.5cm}
   \centering
   \includegraphics[width=0.94\columnwidth]{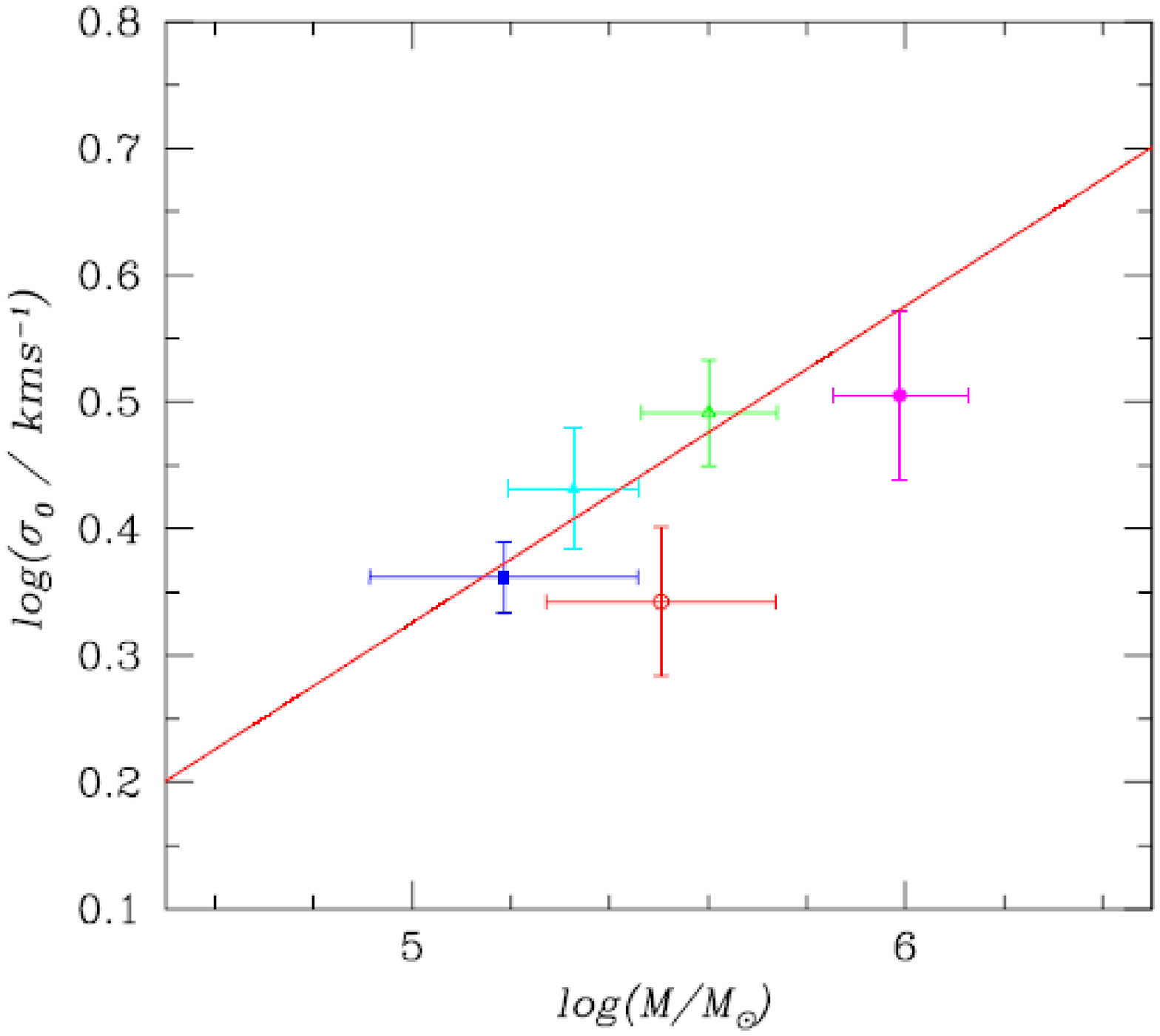}
    \end{minipage}
 \ \hspace{5mm} \
 \begin{minipage}[b]{7.5cm}
  \centering
  \includegraphics[width=0.94\columnwidth]{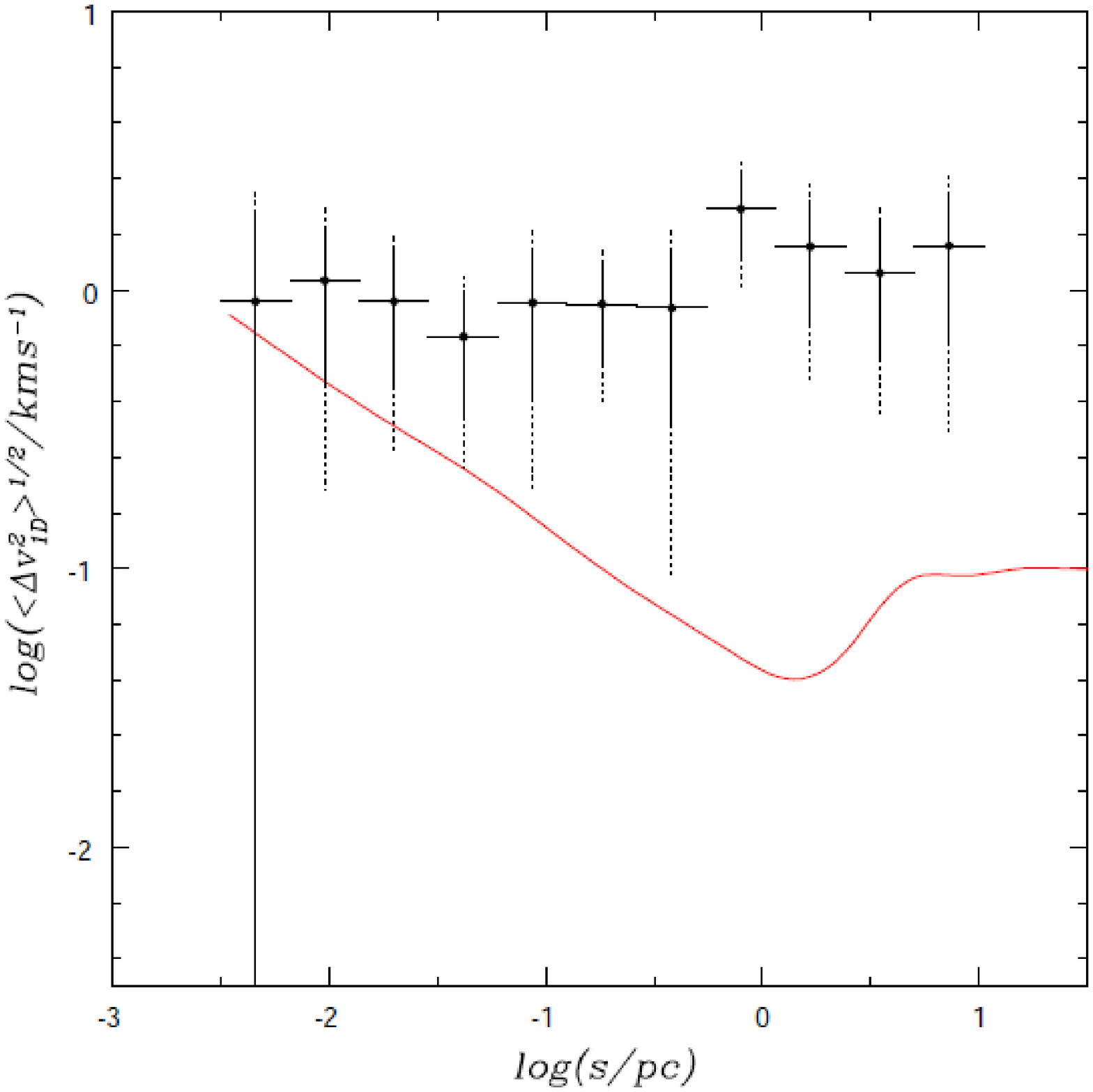}
   \caption{\footnotesize \textbf{a)} Scaling relation for
     Globular Clusters. The solid line is
     the best fit with a $1/4$ slope. \textbf{b)} Solid line: RMS
     One-dimensional, projected relative velocities as function of
     projected separation, from \cite{tremaine:binaries}. Points with
     error-bars: Hipparcos binary catalogue.}
 \end{minipage}
%\caption{Luminosity distance vs redshift for simulated systems.}
\label{fig1}
\end{figure*}

 The outskirts of stellar globular clusters (GCs) are systems which
internal stellar dynamics and structure are well understood without
requiring a dark matter component. On the other hand, recent observations
hint at modified dynamics for the elements in the most external
orbits. The analysis of five GCs by~\cite{hernandez:clusters} has shown
that the dispersion velocity profiles in the external regions, which
experience accelerations $\lesssim a_0$, where \( x \lesssim 1 \), do
not follow the expected Keplerian fall. Instead, a flattening of the
dispersion velocity profile is observed and a scaling law of the kind
described by \eqref{obs:law} is obtained for these systems as shown
in Figure~\ref{fig1}a).  One could argue that the deviation from the
Keplerian prescription is an effect of the tidal heating by the overall
galactic gravitational field. However, it is hard to believe that  a
tidal field varying in each case could reproduce the scaling law shown
in Figure~\ref{fig1}a.

  The most intriguing observable to date concerns the orbits of
wide binary systems, which are ubiquitous in our Solar vicinity
and their binding energy competes with the close encounters with
other stars and with the tidal force of the Galaxy. Recently
\cite{tremaine:binaries} have simulated the evolution of 
\( 5\times10^{4} \) binaries with $ 1 M_{\odot}$ 
in the solar neighbourhood under Newtonian gravity. 
The simulation takes into account the cumulative effects of the
Galactic tidal field, determined observationally through the Oort
constants, and encounters with field stars. The authors compute the
current distribution of separations and relative velocities. The
effective tidal radius for $1 M_{\odot}$ binaries is $ r_{\rm t} = 1.7
\textrm{pc} $. Below this separation, the relative velocity $\Delta V$
in a binary system  should scale with separation $s$ following
Kepler's third law, i.e. $\Delta V \propto r^{-1/2}$. 

  In~\cite{tremaine:binaries}, the RMS relative velocity, projected
along the line of sight, is plotted against the projected separation, as
reproduced in Figure~\ref{fig1}b). The above provides a test for
gravity since the acceleration for particles orbiting a $1 M_{\odot}$
star falls below the transition value $a_0$ at separations $s =
3.4\times 10^{-2} \textrm{pc}$. In the non-Newtonian 
regime of ENG, one should observe relative velocities which do not
decrease with separation. Such trend follows because the binary systems
are much more robust to the tidal forces than under the
Newtonian regime.

  The wide binary systems of the Hipparcos and the
SDSS catalogues have been analysed in \cite{hernandez:binaries}. The
sample selects stars of $\sim 0.5 M_{\odot}$, with no other close
neighbours. When plotted on the projection space of
Figure~\ref{fig1}b), the distribution of stars in each catalogue lies above
the prediction of \cite{tremaine:binaries} even after taking into
account observational uncertainties. Instead, a constant relative
velocity distribution is clear, supporting the non-Newtonian behaviour
predicted by ENG. 
 
  In summary, the introduction of a new fundamental
constant of nature \( a_0 \) yields a general scale-invariant
non-relativistic theory which extends Newtonian gravity and
merges naturally with MOND.  ENG proves useful in explaining
the dynamics of several astrophysical systems, ranging from 
our Solar System to galaxies.  Future observations of the mentioned
environments are required to test this theory in detail. 
\\

\textbf{ACKNOWLEDGEMENTS:} This work was supported by the following
grants: UNAM-DGAPA (PAPIIT IN116210-3, IN103011-3), CONACyT: 207529,
25006, 26344. JCH acknowledges sponsorship from CTIC-UNAM and CONACYT.

%%%%%%%%%%%%%%%%%%%%%%%%%%

\bibliographystyle{plain}

\end{document}